\documentclass[fleqn,twoside]{article}
\usepackage{espcrc2}

\usepackage{graphicx}
\usepackage[figuresright]{rotating}

\newcommand{\AmS}{{\protect\the\textfont2
  A\kern-.1667em\lower.5ex\hbox{M}\kern-.125emS}}

\title{Neutrino Physics around MeV Energies
}

\author{Kunio Inoue\address{Research Center for Neutrino Science, Tohoku University, Sendai, Miyagi 980-8578, Japan}
and Hisakazu Minakata\address{Department of Physics, 
Tokyo Metropolitan University, Hachioji, Tokyo 192-0397, Japan} 
}

\begin{document}

\begin{abstract}

We present a brief informative overview of a broad ranges of subjects,
the solar,  the reactor, the geo and the supernova neutrinos
(but without excluding possible biases), the topics 
which consist of the session
``Neutrino Physics around MeV Energies'' in NOW2006.
Contrary to the naive expectation, the field is found
to be very active in improving the performance of existing detectors,
and in preparation for the development coming in the near future.
The former includes excellent performance of SNO $^3$He detector,
successful reconstruction of SK III, and KamLAND's new ``4$\pi$'' calibration.
The next data release from these experiments will be very exciting.
The latter includes effort for lowering threshold in SK III, KamLAND's
purification by which geo-neutrino observation becomes much cleaner,
and the forthcoming $^7$Be neutrino measurement by BOREXINO
and KamLAND. Possible detection of relic supernova neutrinos and
non-zero effect of $\theta_{13}$ would bring us great excitement.
Furthermore, improved new measurement of heavy element abundance
in the solar atmosphere resulted in a solar model with much lower CNO
$\nu$ flux and with disagreement with helioseismology, thereby bringing 
us a new solar puzzle.

\end{abstract}

\maketitle

\section{Solar neutrinos}

The past and the ongoing solar neutrino experiments cover  
a wide variety of interesting aspects of solar neutrinos \cite{solar}. 
Davis' chlorine experiment which pioneered the solar neutrino search 
has been continuing for more than 30 years. 
Gallium experiments have a sensitivity on the most elemental pp 
neutrinos. 
Super-Kamiokande (SK) is performing precise measurement of mainly 
$^8$B neutrinos with highest statistics. 
SNO has provided an evidence for neutrino flavor transformation with 
CC/NC measurement \cite{maneira_now} .

Missing piece from the experimental side is a realtime spectrum 
measurement at energies below 5 MeV.
Individually measured fluxes of pp, $^7$Be and pep neutrinos will be 
the key to complete the verification of the standard solar model (SSM). 
A measurement of CNO cycle neutrinos would be a realization of 
Davis' initial dream and it will contribute to construct the stellar evolution model. 
Along these motivations, there are two forthcoming experiments; 
KamLAND solar phase \cite{shirai_now} and 
BOREXINO \cite{ranucci_now}.
They will start by observing $^7$Be neutrinos which have high signal rate 
and the characteristic spectrum edge.
Also, pep/CNO neutrinos are seen within their scopes but they require 
special treatment on $^{11}$C spallation background.
This background is more serious for KamLAND which is placed in 
shallower site but tagging the 
associated neutrons ($\sim$95\% BR) will drastically improve the 
situation. A new electronics designed for maximizing the rejection 
efficiency is being developed. 
Proposed SNO$^+$ with liquid scintillator will have much better sensitivity 
to these neutrinos thanks to its size and depth. 
The realtime measurement of pp neutrino is very challenging. 
It requires very powerful background rejection method/apparatus 
such as delayed coincidence or non-carbonic ultra pure detector. 
LENS \cite{grieb_now} revived an idea of indium loaded liquid scintillator 
(3 fold delayed coincidence) with sophisticated fine segmentation techniques.

\begin{table*}[htb]
\caption{Predicted solar neutrino fluxes from
seven solar models. Taken from \cite{bahcall-etal}. 
The table presents the predicted fluxes, in
units of $10^{10}(pp)$, $10^{9}({\rm \, ^7Be})$, $10^{8}(pep, {\rm
^{13}N, ^{15}O})$, $10^{6} ({\rm \, ^8B, ^{17}F})$, and
$10^{3}({\rm hep})$ ${\rm cm^{-2}s^{-1}}$ for the seven different 
solar models calculated in \cite{bahcall-etal}. 
The last two models utilize the new metal abundance data. }
\vspace{0.5cm}
\begin{center}
\begin{tabular}{lcccccccccc}
\hline \hline
Model & pp & pep & hep & $^7$Be &  $^8$B & $^{13}$N & $^{15}$O & $^{17}$F \\
\hline
BP04(Yale) & 5.94 & 1.40 & 7.88 & 4.86 & 5.79 & 5.71 & 5.03 & 5.91 \\
BP04(Garching) & 5.94 & 1.41 & 7.88 & 4.84 & 5.74 & 5.70 & 4.98 & 5.87 \\
BS04 & 5.94 & 1.40 & 7.86 & 4.88 & 5.87 & 5.62 & 4.90 & 6.01 \\
BS05($^{14}$N) & 5.99 & 1.42 & 7.91 & 4.89 & 5.83 & 3.11 & 2.38 & 5.97 \\
BS05(OP)& 5.99 & 1.42 & 7.93 & 4.84 & 5.69 & 3.07 & 2.33 & 5.84 \\
BS05(AGS,OP) & 6.06 & 1.45 & 8.25 & 4.34 & 4.51 & 2.01 & 1.45 & 3.25 \\
BS05(AGS,OPAL) & 6.05 & 1.45 & 8.23 & 4.38 & 4.59 & 2.03 & 1.47 & 3.31 \\
\hline
\end{tabular}
\end{center}
\label{7models}
\end{table*}

Measurement of expected upturn in low energy $^8$B neutrino 
spectrum will possibly provide the still missing 
experimental proof of the LMA solution in the solar neutrino data.
Because the flux is small, it requires a very big detector.
The largest solar neutrino detector, SK-III which has been restarted 
with full PMT installation in July 2006, is aiming at the measurement 
\cite{smy_now}. 

From the theory side or the viewpoint of solar model building, 
recent progresses on the improved cross section measurement with 
LUNA \cite{LUNA_now} and improved modeling of solar atmosphere elements 
(less volatile elements such as C, N, O) provided a big change in 
the prediction of neutrino fluxes \cite{serenelli_now}. 
Less C, N, O abundances correspond to smaller opacity, 
smaller temperature gradient, lower core temperature,   
and consequently result in smaller flexes of 
$^7$Be, $^8$B and CNO cycle neutrinos. 
Estimation of CNO cycle neutrino flux became smaller also by 
the new $S_{1,14}$ measurement and directly by the smaller 
abundances. During the last a few years, the estimated fluxes 
became lower by a factor of 3. See Table~\ref{7models}. 
The problem is that the new model with lower metal abundance 
predicts shallower convection region and the compatibility with 
helioseismology becomes much worse \cite{bahcall-etal}. 
Thus, better determination of heavy element abundances in the 
solar atmosphere has produced a new solar model puzzle \cite{serenelli_now}. 
Importance of improving the accuracy of solar flux measurement is 
still unchanged.

Subdominant effects on solar neutrino deficit other than the LMA 
is still an open question. Various possibilities of time variation
in the solar neutrino data were speculated again in these days. 
An example is a 2.4 $\sigma$ discrepancy of the Ga event rate 
before and after 1998 \cite{pulido_now}. 
It can be explained by the existence of a sizable neutrino magnetic 
moment without causing any inconsistency on the other solar and 
reactor data. 
Examination of this kind of subdominant effects can be on a
task list of future low energy solar neutrino experiments and a 
precision reactor experiment.

\section{Reactor neutrinos}

KamLAND experiment has contributed to solve the solar neutrino 
problem \cite{KamLAND} and it is the only experiment 
which provided the evidence for spectral distortion due to  
the solar $\Delta m^2$ oscillation \cite{KL_evidence}. 
The evidence was the prerequisite for the accurate determination 
of the solar squared-mass-difference that followed. 
Its precision is however still limited by the statistics. 
On the other hand, main limitation to the accuracy of the mixing angle 
determination is coming from the systematic errors of fiducial volume 
and background estimation.
A large fiducial volume error of 4.7\% has been assigned \cite{KL_evidence} 
because a whole-volume calibration with radioactive sources was missing.
KamLAND developed a new positioning device called ``$4\pi$'' 
which suspends a long pole with two wires. 
By controlling the length of the wires and tilting the pole, radioactive 
sources attached at a point of the pole can be located at all positions 
in the detector. Furthermore, the new purification system constructed 
for the KamLAND solar phase will eliminate the dominant background, 
$\rm ^{13}C(\alpha, \it n)$. 
Also a sophisticated analysis using larger (may be entire) fiducial volume 
is developed. All these efforts will contribute to improve the precision of 
oscillation parameter determination with reactor anti-neutrinos and will 
provide a sensitive test of CPT invariance in comparison with global fit 
of solar neutrino data.

The remaining unique unmeasured mixing angle $\theta _{13}$ 
is the target of very active investigations.  
Various experiments and observations are in fact have sensitivities 
to this parameter. 
Examples include long baseline accelerator experiments, reactor $\theta _{13}$ experiments, matter effect of neutrino oscillation in supernovae, 
supernova nucleosynthesis and so on.
Many reactor $\theta _{13}$ experiments have been proposed because 
it is competitive, relatively cheap, and also is complementary with 
accelerator experiments \cite{MSYIS}. 
In consideration of the balance of quickness and sensitivity, 
some projects were terminated and some are suspended. 
Now, apparently they are converged into 3 projects, 
the quickest Double CHOOZ \cite{Cabrera} in France, 
the best sensitivity Daya Bay in China, and RENO in 
Korea \cite{joo_now}. 
They are planned to start gradually from 2008 (CHOOZ far) and 
reach the best sensitivity of $\sin^22\theta_{13} \simeq0.01$ 
around 2013 (Daya Bay).

\section{Geo-neutrinos}

The earth is emitting anti-neutinos that come from radioactive decays 
of uranium-, thorium-series and potassium.
End point energies of the geo-neutrino are 3.27 MeV, 2.25 MeV 
and 1.31 MeV for $^{238}$U-series, $^{232}$Th-series and $^{40}$K,
respectively, while reactor neutrinos extends up to $\sim$10 MeV. 
If one employs the well-established method with reactor neutrinos, 
the inverse beta decay ($\bar{\nu}_e+p\to e^++n$), 
its reaction threshold is 1.8 MeV and neutrinos from $^{40}$K is 
unobservable, leaving it to future experiments with new detection methods.
These radioactive decays are believed to be the major heat source of 
the earth and the bulk silicate earth model predicts about 20 TW heat 
production from them. 
Uncovering the nature of the heat source is very important to understand 
various aspects of earth formation and terrestrial dynamics; 
mantle convection relevant for earthquake and eruption, 
outer core convection driving terrestrial magnetism, etc. 
It will eventually lead to the construction of an observation-based robust 
earth model of formation and evolution.

\begin{figure}[htb]
\begin{center}
\includegraphics[width=18pc]{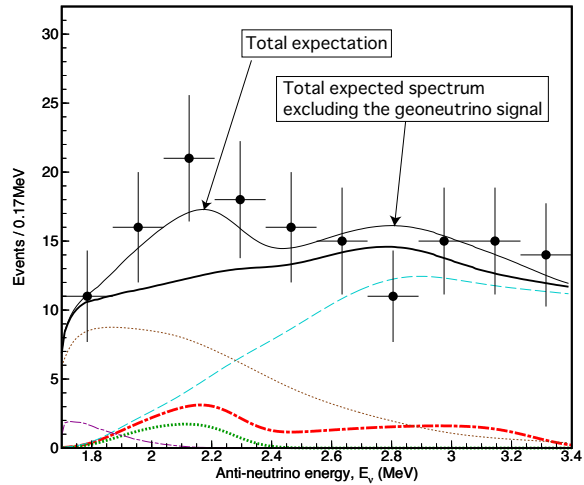}
\vspace{-0.6cm}
\caption{A clear excess of events over the estimated background 
denoted by the thick-solid line is observed in neutrino 
energy range of 1.8$-$3.3 MeV. 
It is strongly indicative of the signal due to geo-neutrinos, and in fact 
it is consistent with the expectation by the bulk silicate earth model, as 
indicated by the thin-solid line. For more details, see 
\cite{kamland_nature}. 
Taken from http://www.awa.tohoku.ac.jp/KamLAND/
GeoNeutrino/index$-$e.html. 
}
\label{KL_geonu}
\end{center}
\end{figure}

Despite geo-neutrino observation is very useful to study the heat 
production in the earth, its detection hasn't been achieved for a 
long time due to the very low interaction rate. 
The size of KamLAND is two orders of magnitude larger than the 
detectors used in the previous reactor experiments and 
it finally made experimental investigation of geo-neutrinos possible. 
KamLAND used anti-neutrino events with neutrino energy of 
above 3.4 MeV to estimate reactor neutrino contamination in 
the geo-neutrino region below 3.4 MeV. Using a tighter selection 
criteria, KamLAND observed 152 events in 750 days of livetime, where 
estimated number of background is $127\pm 13$ 
(dominated by low energy component of reactor neutrinos and 
$\rm ^{13}C(\alpha, \it n)$ from radioactive impurities) \cite{kamland_nature}. 
The number of excess events, $25^{+19}_{-18}$, is consistent with 
bulk silicate earth model prediction, 19 events. 
It is not very significant yet, but KamLAND established the method 
to observe geo-neutrinos and it certainly opened the door to a new branch of 
science, the ``neutrino geophysics''. 

However, it is also clear that future geo-neutrino detectors are 
better to be located far away from nuclear reactors and, if possible, 
to have much larger size. 
It should also be noticed that translation of geo-neutrino event rate 
to the amount of radioactive heat production in the earth requires 
a model of radioactive source distribution.
Therefore, it is highly desirable to have directional information of 
geo-neutrinos. 
A method for directional measurement of anti-neutrinos is under study 
to statistically distinguish between reactor neutrinos and geo-neutrinos from 
the crust and the mantle \cite{shimizu_now}. 
It will merit to improve the sensitivities on solar/geo-reactor anti-neutrinos, 
and may also be useful for advanced warning of supernova direction for 
optical observations, the SNEWS project \cite{SNEWS_now}. 

There are many ways to proceed; 
A possibility is to use geo-neutrino data to verify or reject various models. 
Very sophisticated models have been created based on 
Preliminary Reference Earth Model and a crustal characterization 
with 2 x 2 degree unit for structural details and elemental abundances globally constrained with the BSE model for geochemical details, which lead to site specific predictions for 
geo-neutrinos  \cite{rotunno_now}. 
%
The best thinkable possibilities include, for example, global observatory
network, movable detector and the directional measurement.
Hanohano project is a proposed deep ocean anti-neutrino observatory 
and the detector is designed to be movable \cite{dye_now}. 
If it is placed at around Hawaii, it should be able to probe geo-neutrinos 
from the mantle region in the earth. 
Its contribution is estimated to be as high as $\simeq$75\% because the oceanic 
crust is much thinner and less condensed with uranium and thorium than the continental one.
%
It is also suggested \cite{dye_now} that, because it is movable, 
it may have other physics capabilities such as 
precise solar parameter ($\Delta m^2_{21}$ and $\theta_{12}$) determination, 
$\theta_{13}$ measurement, and even the mass hierarchy determination 
by moving to a preferred site.

\section{Supernova neutrinos}
\label{SNnu}

The dynamics of supernova (SN) explosion is still a 
far-from-understood problem which is under active investigation.  
For its most recent status, see \cite{cardall_now}. 
Though a great amount of efforts have been devoted to do realistic 
simulations of explosion by including various effects like multi-dimensional 
hydrodynamics, an improved neutrino transport, magnetic field, 
and rotation of star etc., the model simulations do not show explosion 
in a robust way. 
Therefore, we definitely need high-statistics data of neutrinos from 
future galactic SN to understand the mechanism of SN explosion.

Are we well prepared for the next supernova?
The answer is, we think, yes and no. 
The answer is yes because SK is now fully 
reconstructed and KamLAND is continuously taking data. 
But, the question is; Do they alive 30$-$50 
years from now, and even so are they enough to acquire 
necessary informations?

\subsection{Supernova core diagnostics by neutrino observation}
\label{diagnostics}

\begin{table}[t]
\caption{Calculated numbers of events expected without and 
with (in parenthesis, assuming the normal hierarchy) neutrino oscillation 
in SK with a 5 MeV threshold with a supernova at 10 kpc, 
which are taken from \cite{beacom-vogel98} and \cite{MN90}, respectively. 
The numbers in KamLAND \cite{KamLAND_US} are given only 
for an alternative channel. 
$\nu_{\mu}$ and $\nu_{\tau}$ are assumed to be identical. 
}
\vspace{0.5cm}
\begin{tabular}{|l|l|}
\hline
reaction in SK & event No. in SK \\
\hline\hline
$\bar{\nu}_e + p \rightarrow e^+ + n$ & 8300 (6500) \\
\hline
$\nu_\mu + ^{16}{\rm O} \rightarrow \nu_\mu + \gamma + X$ & 355 \\
$\bar{\nu}_\mu + ^{16}{\rm O} \rightarrow \bar{\nu}_\mu + \gamma + X$ & \\
\hline
$\nu_e + e^- \rightarrow \nu_e + e^-$ & 200 (250) \\
$\bar{\nu}_e + e^- \rightarrow \bar{\nu}_e + e^-$ & \\
\hline
$\nu_\mu + e^- \rightarrow \nu_\mu + e^-$ & 60 (57) \\
$\bar{\nu}_\mu + e^- \rightarrow \bar{\nu}_\mu + e^-$ & \\
\hline\hline
reaction in KL (not in SK) & No. in KamLAND\\
\hline
$\nu_e + ^{12}\mbox{C} \rightarrow ^{12}\mbox{N}^* + e^-$ & 2 (27) \\
\hline
$\bar{\nu}_e + ^{12}\mbox{C} \rightarrow ^{12}\mbox{B}^* + e^+$ & 7 (7) \\
\hline
$\nu_\mu + ^{12}\mbox{C} \rightarrow ^{12}\mbox{C}^* + \nu_\mu$ & 50 (60) \\
\hline
\end{tabular}
\label{eventsSK-KL}
\vspace{-0.3cm}
\end{table}

In the context of flavor-dependent reconstruction of supernova 
neutrino fluxes the answer to the above question is clearly {\em no}. 
A SN at 10 kpc is believed to produce events in SK and KamLAND as 
tabulated in Table~\ref{eventsSK-KL}. 
The numbers are given only for the normal hierarchy, but it is very 
much dependent of the hierarchy \cite{dighe-smi,MN01}. 
Even if we assume that the mass hierarchy is determined by other means, 
reconstruction of luminosity and spectra of three effective species 
of SN neutrinos, $\nu_{e}$, $\bar{\nu}_{e}$, and $\nu_{\mu} = \nu_{\tau}$, 
does not appear to be possible with the observation by SK and KamLAND only.  
The most difficult part is to obtain spectral information of $\nu_{\mu}$ 
because NC measurement is essentially ``counting the number'' experiment. 
An organized efforts are called for to define a well-thought strategy 
for future SN neutrino observation with practical set up of 
experimental arrays.

\subsection{Relic supernova neutrinos}

\begin{figure}[htb]
\begin{center}
\includegraphics[width=18.5pc]{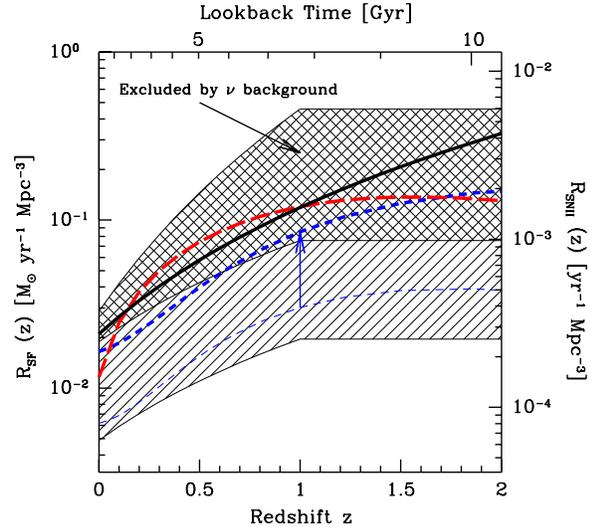}
\vspace{-1.0cm}
\caption{Star formation rate as a function of red shift estimated by 
various methods. Taken from \cite{Strigari2}. 
The entire cross-hatched plus shaded region is consistent with the 
results of the 2dF and SDSS cosmic optical spectrum. 
The upper cross-hatched region is excluded by relic SN $\nu$ observation. 
Three recent results for the star formation rate 
are also shown with long-dashed red line (Dahlen et al.), 
solid black line (GALEX), and short-dashed blue line (Cole et al.). 
The blue up-going arrow indicates the size of dust correction. 
For more details, see \cite{Strigari2}. 
}
\label{SFR}
\end{center}
\end{figure}

The relic SN neutrinos, the ones integrated over the past history of 
universe, will give us a mean for direct probe of the type II supernova rate. 
Then, it can be translated into the constraint on the star formation rate 
in the universe  which has a large uncertainty at the moment. 
For recent discussion, see e.g., \cite{Ando,Fukugita,Strigari,Strigari2} 
and an extensive references in \cite{Strigari2}. 
Figure~\ref{SFR} summarizes the situation. 
Since the prediction by various models is either above or close to 
the limit placed by SK \cite{relic_SK}, there is an exciting possibility 
that it will be detected soon. 
Otherwise, we would need either Gd loading of SK III which is the 
subject of the next subsection, or a megaton class detector. 
See \cite{lunardini} for these points.

\subsection{Gd-loaded Super-Kamiokande}

A proposal of Gd-loaded SK (``GADZOOKS'') that has 
been put forward by Beacom and Vagins \cite{GdSK} is an extremely 
interesting option. 
If realized it will allow us to tag neutrons at $\simeq$90\% efficiency which 
tremendously enhances detectability of SN relic neutrinos as well as 
adding new power to flavor dependent reconstruction of $\nu$ fluxes.
The real problem for this proposal is the experimental issues, 
such as whether one can guarantee that water transparency and 
radio purity can be maintained, and no damage is given to the 
water tank etc., as addressed in the talk in this workshop \cite{vagins}.

As it stands, neutrino physics around MeV energies is a very exciting active field. 
We must mention that while we tried to cover all the topics covered in the 
session but failed to discuss them in sufficient depth. 
Nonetheless, we hope that this manuscript succeeds to convey the 
exciting atmosphere felt by people in the community.

We thank the organizer of NOW2006 for cordial invitation and offering us 
the chance of organizing the MeV neutrino session.  
This work was supported in part by the Grant-in-Aid for Scientific Research, 
No. 16340078, Japan Society for the Promotion of Science.

\end{document}